\documentstyle[12pt]{article}
\input latexsym.sty
\begin{document}

\begin{center}

{\LARGE \bf Bound state problem for Dirac particle \\
in external static charge distribution in $(1+1)$-dimensions}

\vspace{2 cm}

{\large \bf Fuad M. Saradzhev}\footnote{e-mail: physic@lan.ab.az}

\vspace{5 mm}

{\it The Abdus Salam International Centre for Theoretical Physics,\\
Strada Costiera 11, 34014 Trieste, Italy}

\vspace{5 mm}

{\it and}

\vspace{5 mm}

{\it Institute of Physics, Academy of Sciences of Azerbaijan,\\
Huseyn Javid pr. 33, 370143 Baku, Azerbaijan}\footnote{Permanent
address}

\vspace{1 cm}

{\bf Abstract}

\end{center}

\vspace{5 mm}

We study the self-interaction effects for the Dirac particle
moving in an external field created by static charges in $(1+1)$-
dimensions. Assuming that the total electric charge of the
system vanishes, we show that the asymptotically linearly rising
part of the external potential responsible for nonexistence of
bound states in the external field problem without self-interaction
is cancelled by the self-potential of the zero mode of the Dirac
particle charge density. We derive the Dirac equation which includes
the self-potential of the non-zero modes and is nonlinear. We solve
the spectrum problem in the case of two external positive charges
of the same value and prove that the Dirac particle and external
charges are confined in a stable system.

\newpage

\section{Introduction}

The problem of motion of the Dirac particle in an external field
can be studied in different approximations. In the vanishing
self-interaction approximation we neglect the effects of
self-interaction, i.e. of the fact that the Dirac particle creates
its own radiation field and interacts with it. Only the interaction
between the Dirac particle and external field is taken into
account. If the external field is created by static charges, then
this interaction is described by the Coulomb potential.

However, the vanishing self-interaction approximation cannot be
applied universally; in some cases it leads to incorrect results
and even to paradoxes. In particular, if we take in $(1+1)$-
dimensions a system of two spin-$1/2$ particles of opposite
charges coupled by instantaneous Coulomb interaction and assume that
one of the particles is much heavier (proton), while the lighter
particle (electron) moves in the Coulomb field of the heavier one,
then it turns out that the system has not discrete energy levels
[1]. In other words, in $(1+1)$-dimensions the hydrogen atom does
not exist. That happens not only for hydrogenlike systems with an
infinitely heavy source of potential, but also for positroniumlike
systems.

It is therefore of principal importance to include self-interaction.
We need the self-field effects to obtain the full picture of the
interaction between the electromagnetic field and Dirac matter as
well as to make that picture selfconsistent. In the self-field
formulation, the electromagnetic field has as its source all the
charged particles which in turn move in this field. The total
electromagnetic field is a sum of an external and a self-field parts.
The external part is created by some external sources which are not
dynamically relevant, and the self-field is created by the Dirac
particle itself.

In the present paper we aim to study the effects of self-interaction
for the Dirac particle in $(1+1)$-dimensions in connection with the
result of [1]. We want to determine whether in the presence of the
self-field the Dirac particle and external charges can be confined
in a stable system characterized by discrete energy levels.

Models in $(1+1)$-dimensions are interesting as simpler models
for discussion of different aspects of realistic particle physics
models in $(3+1)$-dimensions. At the same time, $(1+1)$-dimensional
models are interesting in their own right and have some pecularities
which make physics in $(1+1)$-dimensions different in principle
from one in $(3+1)$-dimensions. One of these pecularities is that
the Coulomb potential on line is linearly rising at spatial
infinities. Just the linear Coulomb potential is responsible for
the paradox mentioned above.

One-dimensional models of spin-$1/2$ particles have applications in
condensed matter physics, too. It is enough to mention the
one-dimensional model of electronic liquid or the Thirring model.
The problem of existence of bound states for a Dirac particle
in the presence of a static charge distribution may be useful
for understanding of formation and decay of bound states for
these particles in other one-dimensional models with more
complicated interaction.

Our paper is organized as follows. In section 2, we neglect first
self-interaction and consider the Dirac particle moving in a
potential created by external static charges. We find asymptotics
of the eigenfunctions of the Dirac Hamiltonian at spatial
infinities. In accordance with [1], for any value of energy the
eigenfunctions are not normalizable and cannot correspond to a
discrete spectrum. In section 3, we introduce the self-field of
the Dirac particle. Following the method of [2-4], we derive the
Dirac equation which includes the nonlinear self-field term.
In section 4, we study this equation for a specific choice of
the external potential, namely for the case of two external
positive charges of the same value. We solve the spectrum
problem in the approximation when the discrete energy values are
determined by the interaction between the Dirac particle and external
field, while the self-interaction shifts these values by a small
amount. Section 5 contains our conclusions.

\newcommand{\ren}{\renewcommand{\theequation}{2.\arabic{equation}}}
\newcommand{\new}[1]{\renewcommand{\theequation}{2.\arabic{equation}#1}}
\newcommand{\add}{\addtocounter{equation}{-1}}

\section{Dirac particle in external field}

The Dirac equation for a particle of charge $e$ and mass $m$
moving in an external field is
\ren
\begin{equation}
[{\gamma}^{\mu} (i{\hbar}c {\partial}_{\mu} - e A_{\mu}^{ext}) 
- mc^2] {\psi}(x) =0 ,
\label{eq: dvaodin}
\end{equation}
where $({\mu},{\nu} = \overline{0,1})$, ${\gamma}^{\mu}$ are $2 \times 2$
Dirac matrices, 
\begin{displaymath}
{\gamma}^0 =
\left( \begin{array}{cc}
0 & 1\\
1 & 0 
\end{array} \right),
\;\;\;\;\;\;
{\gamma}^1 =
\left( \begin{array}{cc}
0 & -1 \\
1 & 0
\end{array} \right),
\end{displaymath}
and the field ${\psi}(x) = {\psi}(x_1,t)$ is two-component
Dirac spinor. The partial derivatives are defined as ${\partial}_0 =
{\partial}/c{\partial}t$, ${\partial}_1 = {\partial}/{\partial}x_1$.

We assume that the external field is defined by
\[
A_{\mu}^{ext}(x_1,t) = (A_0^{ext}=A_0^{ext}(x_1) , A_1^{ext}=0),
\]
which is a time-independent scalar potential. We assume next that the
potential is created by static charges, so we can use the following
ansatz
\ren
\begin{equation}
A_0^{ext}(x_1)=- \frac{1}{2} Q^{ext} |x_1| + \bar{A}_0^{ext}(x_1),
\label{eq: dvadva}
\end{equation}
where $Q^{ext}$ is total external charge, and $\bar{A}_0^{ext}$ is
a potential which does not increase at spatial infinities, i.e.
$\bar{A}_0^{ext}(x_1 = \pm \infty)$ are finite constants.

In particular, one external charge $q$ at the point $x_1=0$
creates the current $j_0^{ext}=q\delta(x_1)$ that corresponds
to the scalar potential $A_0^{ext} = -\frac{1}{2} q|x_1|$.
Two external charges of the same value $q$ taken at the points
$x_1=a$ and $x_1=-a$ create the current $j_0^{ext}=
q (\delta(x_1-a) + \delta(x_1+a))$ and potential of the form (2.2)
with $Q^{ext}=2q$ and
\ren
\begin{equation}
\bar{A}_0^{ext} =
\left\{ \begin{array}{cc}
q(|x_1|-a) & {\rm for} \hspace{5 mm} |x_1| \leq a,\\
\vspace{2 mm} \\
0 & {\rm for} \hspace{5 mm} |x_1| \geq a.
\end{array}
\right.
\label{eq: dvatri}
\end{equation}
We can easily check that the ansatz (2.2) is valid for an arbitrary
number of external static charges $q_1,q_2,q_3,...$ . Therefore,
the potential for these charges can be always given as the sum of two
parts, linearly rising and finite at spatial infinities.

If we act on both sides of (2.1) by $[-{\gamma}^{\mu}(i{\hbar}c
{\partial}_{\mu} - eA_{\mu}^{ext}) - mc^2]$, then we come
to the second-order differential equation for $\psi$ [5,6]
\ren
\begin{equation}
\left[ D_{\mu}D^{\mu} + e S^{{\lambda}{\mu}} F_{{\lambda}{\mu}}^{ext}
+ \frac{m^2c^2}{{\hbar}^2} \right] {\psi} =0,
\label{eq: dvacet}
\end{equation}
where
\[
D_{\mu} \equiv {\partial}_{\mu} + i \frac{e}{{\hbar}c} A_{\mu}^{ext}.
\]
This equation looks like the Klein-Gordon one, except the additional
term $eS^{{\lambda}{\mu}}F_{{\lambda}{\mu}}^{ext}$ with
\[
S^{{\lambda}{\mu}} \equiv \frac{1}{2} i [{\gamma}^{\lambda},{\gamma}^{\mu}]_{-},
\]
\[
F_{{\lambda}{\mu}}^{ext} \equiv {\partial}_{\lambda} A_{\mu}^{ext}
- {\partial}_{\mu} A_{\lambda}^{ext}.
\]
In $(3+1)$-dimensions, the spatial components of $S^{{\lambda}{\mu}}$
are related to spin of the Dirac particle, so the additional term describes
interaction between the particle spin and the electromagnetic field.

In $(1+1)$-dimensions, $S^{{\lambda}{\mu}}$ has no spatial components, and
we can not therefore introduce spin. The only nonvanishing component
$S^{01}=i{\alpha}$, where ${\alpha} \equiv {\gamma}^5={\gamma}^0 {\gamma}^1$,
allows us to introduce chirality. If we write ${\psi}$ in the component
form as
\begin{displaymath}
{\psi} =
\left( \begin{array}{c}
u_1 \\
u_2
\end{array} \right)
\end{displaymath}
and define the operators
\[
P_{\pm} \equiv \frac{1}{2} (1 \pm {\alpha}),
\]
then
\begin{displaymath}
P_{+}{\psi} =
\left( \begin{array}{c}
u_1\\
0
\end{array} \right),
%\end{displaymath}
%\hspace{1 cm}
%\begin{displaymath}
\;\;\;\;\;\;
P_{-}{\psi} =
\left( \begin{array}{c}
0\\
u_2
\end{array} \right),
\end{displaymath}
i.e. the upper component is of positive chirality and the lower one -
of negative chirality.

Moreover, there is no magnetic field in $(1+1)$-dimensions, so the only
nonvanishing component of $F_{{\lambda}{\mu}}^{ext}$ is
\[
F_{01}^{ext} = - \frac{{\partial}A_0^{ext}}{{\partial}x_1}
\equiv {\cal E}^{ext},
\]
where ${\cal E}^{ext}$ is the external electric field. The additional term
in (4) becomes
\[
eS^{{\lambda}{\mu}}F_{{\lambda}{\mu}}^{ext}=ie{\alpha}{\cal E}^{ext},
\]
indicating that positive and negative chirality components are coupled
differently to the external electric field.

In the Hamiltonian form the equation (2.1) reads
\ren
\begin{equation}
H{\psi} \equiv i{\hbar}\frac{\partial}{{\partial}t} {\psi} =
\left( i{\hbar}c{\alpha} \frac{\partial}{{\partial}x_1} + {\beta} mc^2
+ e A_0^{ext} \right) {\psi},
\label{eq: dvapet}
\end{equation}
where ${\beta} \equiv {\gamma}^0$. The eigenvalue problem for the
Dirac Hamiltonian, $H{\psi}=E{\psi}$, reduces to the problem of
solving the system of two equations
\new{a}
\begin{equation}
\left( i{\hbar} \frac{\partial}{{\partial}x_1} + \frac{eA_0^{ext} - E}{c} \right)
u_1 = - mcu_2,
\end{equation}
\add
\new{b}
\begin{equation}
\left( -i{\hbar} \frac{\partial}{{\partial}x_1} + \frac{eA_0^{ext} - E}{c} \right)
u_2 =  -mcu_1.
\end{equation}
We easily decouple $u_1$ and $u_2$ and rewrite these equations
equivalently as
\ren
\begin{equation}
\left[ {\hbar}^2 \frac{{\partial}^2}{{\partial}x_1^2} \pm i \frac{e{\hbar}}{c}
{\cal E}^{ext} + \left(\frac{eA_0^{ext}-E}{c}\right)^2 \right]
u_{1(2)}=m^2c^2u_{1(2)},
\label{eq: dvasem}
\end{equation}
the sign $(+)$ corresponding to the positive chirality component, while
$(-)$ to the negative one.

If the Dirac particle and external charges are confined in a stable
system, then equations (2.7) must reveal a set of bound states. 
For a discrete set of energies in the band $|E|<mc^2$, these
equations must have solutions which decrease exponentially at infinities
and are normalizable.

However, for $Q^{ext} \neq 0$, the external potential is asymptotically
linearly rising at spatial infinities, so the term 
$(e^2(Q^{ext})^2x_1^2)/(4c^2)$ dominates in the equations (2.7) and
prevents any bound state. Indeed, the asymptotics of $u_{1(2)}$ for all
possible energies in the band are determined by the equation
\ren
\begin{equation}
\left[ \frac{{\partial}^2}{{\partial}x_1^2} + \frac{e^2(Q^{ext})^2}{4c^2{\hbar}^2}
x_1^2 \right] u_{1(2)} (|x_1| \to \infty) =0.
\label{eq: dvavosem}
\end{equation}
This is the inverted oscillator equation [7]. Its general solution can
be expressed in terms of parabolic cylinder  functions. One way to choose
two linearly independent solutions of (2.8) is to take the real functions
$W(0,x_1)$ and $W(0,-x_1)$ (we follow the notations of [7]) whose
asymptotic behaviour is well-known:
\begin{eqnarray*}
W(0, x_1 \to +\infty) & \approx & \frac{1}{\sqrt{x_1}}
\cos\left(\frac{|eQ^{ext}|}{c{\hbar}} x_1 + \frac{\pi}{4} \right),\\
W(0, x_1 \to -\infty) & \approx & \frac{1}{\sqrt{|x_1|}}
\sin\left(\frac{|eQ^{ext}|}{c{\hbar}} x_1 + \frac{\pi}{4} \right).
\end{eqnarray*}

We can represent $u_{1(2)}(|x_1| \to \infty)$ as linear combinations
of $W(0, x_1 \to \pm \infty)$ with arbitrary coefficients. Regardless
of the choice for these coefficients, the normalization integral
\[
\int^{x_1} dx_1^{\prime} (|u_1(x_1^{\prime})|^2 
+ |u_2(x_1^{\prime})|^2)
\]
diverges for large $x_1$. This means that the probability of finding
the Dirac particle and external charges infinitely separated remains
finite at all times.
Consequently, the equations (2.7) have solutions for an
arbitrary energy, but these solutions can never represent a bound state.

\renewcommand{\ren}{\renewcommand{\theequation}{3.\arabic{equation}}}
\renewcommand{\add}{\addtocounter{equation}{-1}}
\renewcommand{\new}[1]{\renewcommand{\theequation}{3.\arabic{equation}#1}}
\newcommand{\set}{\setcounter{equation}{0}}

\section{Self-interaction}

\set

Let us now introduce the self-field of the Dirac particle and
investigate how it influences the bound state spectrum. To derive an equation
for the Dirac particle in the presence of both the external and 
its own radiation fields , we start first with the action
\[
S= \int_{-\infty}^{\infty} dt \int_{-\infty}^{\infty} dx_1 
[ \bar{\psi}(x_1,t) ({\gamma}^{\mu} i{\hbar}c {\partial}_{\mu} -mc^2)
{\psi}(x_1,t) - j_{\mu}^{tot}(x_1,t) A^{\mu}(x_1,t)
\]
\ren
\begin{equation}
- \frac{1}{4} F_{{\lambda}{\mu}}(x_1,t) F^{{\lambda}{\mu}}(x_1,t) ],
\label{eq: triodin}
\end{equation}
where
\[
j_{\mu}^{tot} = j_{\mu} + j_{\mu}^{ext}
\]
is a total current which includes both the Dirac matter current
$j^{\mu} \equiv e\bar{\psi} {\gamma}^{\mu} {\psi}$ and the one
created by external static charges.

If we expand the field $\psi$ into the Fourier integral
\[
{\psi}(x_1,t)=\frac{1}{\sqrt{2\pi}} \int_{-\infty}^{\infty} dp
\cdot e^{\frac{i}{\hbar}px_1} {\psi}(p,t),
\]
then the charge density of the Dirac matter becomes
\ren
\begin{equation}
j_0(x_1,t)=e{\psi}^{\star}(x_1,t) {\psi}(x_1,t) =
\frac{e}{2\pi} \int_{-\infty}^{\infty} dk \cdot e^{-\frac{i}{\hbar} kx_1}
{\rho}(k,t) ,
\label{eq: tridva}
\end{equation}
where
\[
{\rho}(k,t) \equiv \int_{-\infty}^{\infty} dp {\psi}^{\star}(p+k,t)
{\psi}(p,t).
\]
The zero momentum component of the density determines the matter
charge
\[
Q \equiv \int_{-\infty}^{\infty} dx j_0(x_1,t) = e{\hbar} {\rho}(0,t).
\]
Separating in (3.2) ${\rho}(0,t)$ and ${\rho}(k,t)$ with the non-zero
momentums $k \neq 0$, we rewrite (3.2) as
\ren
\begin{equation}
j_0(x_1,t) =Q\delta(x_1) + \bar{j}_0(x_1,t),
\label{eq: tritri}
\end{equation}
the density
\[
\bar{j}_0(x_1,t) \equiv \frac{e}{2\pi} \int_{-\infty}^{\infty} dk
\cdot e^{-\frac{i}{\hbar}kx_1} \bar{\rho}(k,t),
\]
\[
\bar{\rho}(k,t) \equiv {\rho}(k,t) - {\rho}(0,t),
\]
corresponding  to a zero charge, $\int_{-\infty}^{\infty} dx_1
\bar{j}_0(x_1,t)=0$.

With the choice of the gauge ${\partial}_{\mu} A^{\mu}=0$ , the Maxwell
equations take the form
\ren
\begin{equation}
\Box  A_{\mu} = j_{\mu}^{tot},
\label{eq: tricet}
\end{equation}
and are solved by 
\[
A_{\mu}(x_1,t) =c\int_{-\infty}^{\infty} dx_1^{\prime} \int_{-\infty}
^{\infty} dt^{\prime} D^c(x_1-x_1^{\prime};t-t^{\prime})
j_{\mu}^{tot}(x_1^{\prime},t^{\prime}) 
\]
\ren
\begin{equation}
=A_{\mu}^{self}(x_1,t) + A_{\mu}^{ext}(x_1,t),
\label{eq: tripet}
\end{equation}
where $D^c(x_1,t)$ is the causal Green's function
\[
\Box D^c(x_1,t)=\delta(x_1) \delta(ct),
\]
\ren
\begin{equation}
D^c(x_1,t) \equiv -\frac{1}{(2\pi)^2} \int_{-\infty}^{\infty}
dq_0 \int_{-\infty}^{\infty} dq_1 \frac{1}{q_0^2 - q_1^2 +
i \epsilon} e^{-\frac{i}{\hbar}cq_0t} e^{\frac{i}{\hbar}q_1x_1}
\label{eq: trishest}
\end{equation}
and $\Box \equiv {\partial}_{\mu} {\partial}^{\mu}$.

Substituting (3.3) and (3.6) into the expression for the
electromagnetic field, we get its external and self-field parts
in the form
\[
A_{\mu}^{ext}(x_1,t) = \left( -\frac{1}{2} Q^{ext}|x_1| +
\bar{A}_0^{ext}(x_1) \right) {\delta}_{{\mu}0},
\]
that agrees with (2.2), and
\[
A_{\mu}^{self}(x_1,t) = -\frac{1}{2} Q|x_1| {\delta}_{{\mu}0}
+ \bar{A}_{\mu}^{self}(x_1,t),
\]
with
\[
\bar{A}_{\mu}^{self}(x_1,t) \equiv c\int_{-\infty}^{\infty} dx_1^{\prime}
\int_{-\infty}^{\infty} dt^{\prime} D^c(x_1-x_1^{\prime};t-t^{\prime})
\bar{j}_{\mu}(x_1^{\prime},t^{\prime}),
\]
and
\[
\bar{j}_{\mu}(x_1,t) \equiv j_{\mu}(x_1,t) - Q\delta_{{\mu}0}
\delta(x_1) .
\]
 
Putting both parts together, we see that if the total charge of the
system vanishes
\ren
\begin{equation}
Q^{tot} \equiv Q + Q^{ext} =0,
\label{eq: trisem}
\end{equation}
then the linearly rising potentials produced by the Dirac particle
and external charges cancel each other, and for the total
electromagnetic field we get
\ren
\begin{equation}
A_{\mu}(x_1,t) = \bar{A}_{\mu}^{self}(x_1,t) + \bar{A}_0^{ext}(x_1)
\delta_{{\mu}0}.
\label{eq: trivosem}
\end{equation}

One of the Maxwell equations (3.4) is the Gauss' law
\ren
\begin{equation}
\frac{\partial}{{\partial}x_1} {\cal E} =j_0^{tot} ,
\label{eq: tridevet}
\end{equation}
where the electric field also consists of two parts, self-field and
external,
\[
{\cal E} = F_{01} = {\cal E}^{self} + {\cal E}^{ext}.
\]
Integrating the equation (3.9) over $x_1$, we obtain
\[
{\cal E}^{self}(+\infty) - {\cal E}^{self}(-\infty) =
- \left( {\cal E}^{ext}(+\infty) - {\cal E}^{ext}(-\infty) \right),
\]
i.e. the difference in the values of the self-field electric field
at the ends of $x_1$-line balances the one of the external electric
field, and for the total electric field ${\cal E}(+\infty)=
{\cal E}(-\infty)$.

The vanishing of the total charge allows therefore to balance the
sources of electric flux and is important from a physical point
of view. For $Q^{ext} \neq 0$, the balance is destroyed, and,
as we have seen in the previous section, the Dirac particle
and external charges escape from each other and can not create
a stable system.

The condition (3.7) has its analogues in the second-quantized
version of different $(1+1)$-dimensional models. In the Schwinger
model [8], the total electric charge is known to be zero  on the
physical states [9,10]. Because of Schwinger [8], when a charge
is inserted into the vacuum, the accompanying electric field
polarizes the vacuum producing complete compensation of the
charge. In $(1+1)$-dimensional QCD, we restrict ourselves to color
neutral states, since the presence of uncompensated color charge
in space leads to a growth of the fields at infinities and makes
the total energy of the system infinite [11].

Inserting (3.8) into the action and using a partial integration,
we have
\[
S = S_0 + S_{self},
\]
\[
S_0 \equiv \int_{-\infty}^{\infty} dt \int_{-\infty}^{\infty} dx_1
[ \bar{\psi}(x_1,t) ({\gamma}^{\mu} i{\hbar}c {\partial}_{\mu}
- mc^2) {\psi}(x_1,t) - e\bar{\psi}(x_1,t) {\gamma}^{\mu}
{\psi}(x_1,t) \cdot \bar{A}_{\mu}^{ext}
\]
\ren
\begin{equation}
- \frac{1}{2} j_{ext}^{\mu} \bar{A}_{\mu}^{ext} ],
\label{eq: trideset}
\end{equation}
\[
S_{self} \equiv \int_{- \infty}^{\infty} dt \int_{- \infty}^{\infty}
dx_1 [ -e\bar{\psi}(x_1,t) {\gamma}^{\mu} {\psi}(x_1,t)
\bar{A}_{\mu}^{self} - \frac{1}{4} \bar{F}_{{\lambda}{\mu}}^{self}
\bar{F}^{{\lambda}{\mu}}_{self} ]
\]
with $\bar{F}_{{\lambda}{\mu}}^{self} \equiv {\partial}_{\lambda}
\bar{A}_{\mu}^{self}
- {\partial}_{\mu} \bar{A}_{\lambda}^{self}$ and $\bar{A}_{\mu}^{ext} \equiv
\bar{A}_0^{ext} {\delta}_{{\mu}0}$. Variation of this action with respect
to the Dirac field yields the following Dirac equation
\ren
\begin{equation}
\left[ {\gamma}^{\mu} (i{\hbar}c {\partial}_{\mu} -e\bar{A}_{\mu}^{ext})
-mc^2 \right] {\psi}(x) = e {\gamma}^{\mu} \bar{A}_{\mu}^{self} {\psi}(x).
\label{eq: triodinodin}
\end{equation}
It is essential that neither the linearly rising part of the external
electric field nor the one of the self-field enter this equation.

Since $\bar{A}_{\mu}^{self}$ is expressed in terms of the current
$\bar{j}_{\mu}$, (3.11) is a non-linear integral equation for ${\psi}$.
In the next section, we will continue our study of the equation
(3.11) for a specific choice of the external electric field.

\renewcommand{\ren}{\renewcommand{\theequation}{4.\arabic{equation}}}
\renewcommand{\add}{\addtocounter{equation}{-1}}
\renewcommand{\new}[1]{\renewcommand{\theequation}{4.\arabic{equation}#1}}
\renewcommand{\set}{\setcounter{equation}{0}}

\section{Example}

\set

{\bf 1.} As an example, let us consider in detail the case of two external
charges of the
same value $q>0$ at the points $x_1=\pm a$ and find first the spectrum
for the vanishing $\bar{A}_{\mu}^{self}$.

With the external potential $\bar{A}_0^{ext}$ given by (2.3) (see Fig. 1.),
the Dirac Hamiltonian is invariant under a modified parity transformation
generated by
\ren
\begin{equation}
\hat{T} \equiv {\beta} \hat{I},
\label{eq: cetodin}
\end{equation}
where the action of $\hat{I}$ on any function of $x_1$ is defined as
\[
\hat{I} f(x_1) = f(-x_1).
\]
The Hamiltonian eigenfunctions must be therefore even or odd under
parity reversal. Since the transformation for ${\psi}$ is
\[
{\psi}(x_1,t) \to {\psi}^{\prime}(x_1,t) = {\beta}
{\psi}(-x_1,t),
\]
we get
\[
u_1(x_1,t) = u_2(-x_1,t)
\]
for even eigenfunctions, and
\[
u_1(x_1,t) = - u_2(-x_1,t)
\]
for odd ones.

To solve the system (2.6a-b) with $\bar{A}_0^{ext}$, we have to find
general solutions in the regions $-a \leq x_1 \leq a$, $x_1 \geq a$ and
$x_1 \leq -a$, and match them at the boundaries so that the eigenfunction
and its first derivative is continuous. We want also to distinguish
the regions of positive and negative values of $x_1$, and so introduce
\begin{eqnarray*}
u_{1(2)}^{+} & \equiv & u_{1(2)}(x_1>0),\\
u_{1(2)}^{-} & \equiv & u_{1(2)}(x_1<0).
\end{eqnarray*}
To simplify the notation, we define in the region $0 \leq x_1 \leq a$
the new dimensionless variable
\ren
\begin{equation}
z_{+} \equiv e^{-i\frac{\pi}{4}} \sqrt{\frac{eq}{{\hbar}c}}
(x_1 - a - \frac{E}{eq} ).
\label{eq: cetdva}
\end{equation}
The system (2.6a-b) then becomes
\new{a}
\begin{equation}
\left( \frac{\partial}{{\partial}z_{+}} + z_{+} \right) u_1^{+}=
i \sqrt{\frac{2}{\Delta}} e^{i\frac{\pi}{4}} u_2^{+},
\end{equation}
\add
\new{b}
\begin{equation}
\left( - \frac{\partial}{{\partial}z_{+}} + z_{+} \right) u_2^{+} =
i \sqrt{\frac{2}{\Delta}} e^{i\frac{\pi}{4}} u_1^{+},
\end{equation}
where $\Delta \equiv (2eq{\hbar})/(m^2c^3)$ is a dimensionless constant.
Decoupling $u_1^{+}$ and $u_2^{+}$, we obtain the system of two
second-order differential equations
\ren
\begin{equation}
\frac{{\partial}^2}{{\partial}z_{+}^2} u_{1(2)}^{+} +
\left[ (2{\nu} \pm 1) - z_{+}^2 \right] u_{1(2)}^{+} =0
\label{eq: cetcet}
\end{equation}
with ${\nu} \equiv - i/{\Delta}$.

The system (4.4) is solved by hypergeometric functions. If we introduce 
\[
y_{+} \equiv z_{+}^2,
\]
\[
v_{1(2)}^{+} \equiv e^{\frac{1}{2}z_{+}^2} \cdot u_{1(2)}^{+},
\]
then (4.4) reduces to
\new{a}
\begin{equation}
y_{+} \frac{{\partial}^2}{{\partial}y_{+}^2} v_{1,+} +
(\frac{1}{2} - y_{+}) \frac{\partial}{{\partial}y_{+}} v_{1,+}
+ \frac{\nu}{2} v_{1,+} =0,
\end{equation}
\add
\new{b}
\begin{equation}
y_{+} \frac{{\partial}^2}{{\partial}y_{+}^2} v_{2,+} +
(\frac{1}{2} - y_{+}) \frac{\partial}{{\partial}y_{+}} v_{2,+}
+ \frac{\nu -1}{2} v_{2,+} =0,
\end{equation}
which are just the hypergeometric function equations. For the first
equation, the linearly independent solutions are $F(-\frac{\nu}{2} ;
\frac{1}{2} ; y_{+})$ and $y_{+}^{1/2}F(\frac{1-\nu}{2} ; \frac{3}{2};y_{+})$,
while for the second one $F(\frac{1-\nu}{2} ; \frac{1}{2}; y_{+})$
and $y_{+}^{1/2}F(1-\frac{\nu}{2};\frac{3}{2};y_{+})$.

A similar analysis can be performed in the region $-a \leq x_1 \leq 0$.
With the variable
\[
z_{-} \equiv e^{-i\frac{\pi}{4}} \sqrt{\frac{eq}{{\hbar}c}} (x_1 + a +
\frac{E}{eq}),
\]
the equations for $u_{1(2)}^{-}$ read
\ren
\begin{equation}
\frac{{\partial}^2}{{\partial}z_{-}^2} u_{1(2)}^{-} +
\left[ (2\nu \mp 1) - z_{-}^2 \right] u_{1(2)}^{-} =0.
\label{eq: cetshest}
\end{equation}
The linearly independent solutions for $v_{1(2)}^{-} \equiv
e^{\frac{1}{2} z_{-}^2} \cdot u_{1(2)}^{-}$ are $F(\frac{1-\nu}{2};
\frac{1}{2}; y_{-})$, $y_{-}^{1/2}F(1-\frac{\nu}{2};\frac{3}{2};y_{-})$
and $F(-\frac{\nu}{2};\frac{1}{2};y_{-})$, $y_{-}^{1/2}F(\frac{1-\nu}{2};
\frac{3}{2};y_{-})$, correspondingly, where $y_{-} \equiv z_{-}^2$.

Taking linear combinations of these solutions, we can construct
eigenfunctions
\begin{displaymath}
{\psi}^{\pm} \equiv
\left( \begin{array}{c}
u_1^{\pm} \\
u_2^{\pm}
\end{array} \right)
\end{displaymath}
which fulfil the matching conditions
\begin{eqnarray*}
{\psi}^{\pm}(x_1=+0) & = & {\psi}^{-}(x_1=-0),\\
\frac{{\partial}{\psi}^{+}}{{\partial}x_1} (x_1 = +0) & = &
\frac{{\partial}{\psi}^{-}}{{\partial}x_1} (x_1=-0),
\end{eqnarray*}
and are even or odd under parity reversal.

The even eigenfunctions up to a constant factor are
\new{a}
\begin{equation}
{\psi}_{even}^{+} = e^{-\frac{1}{2} z_{+}^2}
\left( \begin{array}{c}
\sqrt{\frac{\Delta}{2}} e^{i\frac{\pi}{4}} F(-\frac{\nu}{2};\frac{1}{2};z_{+}^2)
+ G z_{+} F(\frac{1-\nu}{2};\frac{3}{2};z_{+}^2) \\
z_{+} F(1-\frac{\nu}{2};\frac{3}{2};z_{+}^2) - G \sqrt{\frac{\Delta}{2}}
e^{i\frac{\pi}{4}} F(\frac{1-\nu}{2} ; \frac{1}{2} ; z_{+}^2)
\end{array} \right)
\end{equation}
for $0 \leq x_1 \leq a$, and
\add
\new{b}
\begin{equation}
{\psi}_{even}^{-} = e^{-\frac{1}{2} z_{-}^2}
\left( \begin{array}{c}
-z_{-} F(1-\frac{\nu}{2};\frac{3}{2};z_{-}^2) - G \sqrt{\frac{\Delta}{2}}
e^{i\frac{\pi}{4}} F(\frac{1-\nu}{2};\frac{1}{2};z_{-}^2) \\
\sqrt{\frac{\Delta}{2}} e^{i\frac{\pi}{4}} F(-\frac{\nu}{2};\frac{1}{2};
z_{-}^2) - G z_{-} F(\frac{1-\nu}{2};\frac{3}{2};z_{-}^2)
\end{array} \right)
\end{equation}
for $-a \leq x_1 \leq 0$. The constant
\[
G = G({\Delta},z_a) \equiv
\frac{2{\nu}z_a F(1-\frac{\nu}{2};\frac{3}{2};2{\nu}z_a^2) +
F(-\frac{\nu}{2};\frac{1}{2};2{\nu}z_a^2)}
{2{\nu}z_a F(\frac{1-\nu}{2};\frac{3}{2};2{\nu}z_a^2) -
F(\frac{1-\nu}{2};\frac{1}{2};2{\nu}z_a^2)}
\]
is modulo $1$, $|G|^2=1$, and $z_a \equiv (E+eqa)/(mc^2)$.

The odd eigenfunctions are
\new{a}
\begin{equation}
{\psi}_{odd}^{+} = e^{-\frac{1}{2} z_{+}^2}
\left( \begin{array}{c}
\sqrt{\frac{\Delta}{2}} e^{i\frac{\pi}{4}} F(-\frac{\nu}{2};
\frac{1}{2};z_{+}^2) + \bar{G} z_{+} F(\frac{1-\nu}{2};\frac{3}{2};z_{+}^2) \\
z_{+} F(1-\frac{\nu}{2};\frac{3}{2};z_{+}^2) - \bar{G} \sqrt{\frac{\Delta}{2}}
e^{i\frac{\pi}{4}} F(\frac{1-\nu}{2};\frac{3}{2};z_{+}^2)
\end{array} \right)
\end{equation}
for $0 \leq x_1 \leq a$, and
\add
\new{b}
\begin{equation}
{\psi}_{odd}^{-} = e^{-\frac{1}{2} z_{-}^2}
\left( \begin{array}{c}
z_{-} F(1-\frac{\nu}{2};\frac{3}{2};z_{-}^2) + \bar{G} \sqrt{\frac{\Delta}
{2}} e^{i\frac{\pi}{4}} F(\frac{1-\nu}{2};\frac{1}{2};z_{-}^2) \\
- \sqrt{\frac{\Delta}{2}} e^{i\frac{\pi}{4}} F(-\frac{\nu}{2};
\frac{1}{2};z_{-}^2) + \bar{G} z_{-} F(\frac{1-\nu}{2};\frac{3}{2};z_{-}^2)
\end{array} \right)
\end{equation}
for $-a \leq x_1 \leq 0$, where
\[
\bar{G}({\Delta},z_a) = - G({\Delta},-z_a).
\]

For $x_1 \geq a$ it can be checked that
\ren
\begin{equation}
{\psi}^{+}_{even} = s
\left( \begin{array}{c}
E - i {\kappa}\\
mc^2
\end{array} \right)
e^{- \frac{1}{{\hbar}c} {\kappa} x_1}
\end{equation}
with
\[
{\kappa} \equiv \sqrt{m^2c^4 - E^2}
\]
satisfies the Dirac equation. Matching (4.7a) and (4.9) at
$x_1=a$ and eliminating $s$ we obtain the equation that determines the
spectrum of the even bound states
\ren
\begin{equation}
E = mc^2 \cos\left( {\lambda}_{G}(E) \right)
\end{equation}
where
\[
{\lambda}_{G} \equiv \arg \left( \frac{2{\nu} G z_0 F(\frac{1-{\nu}}{2};
\frac{3}{2};2{\nu}z_0^2) - F(-\frac{\nu}{2};\frac{1}{2};2{\nu}z_0^2)}
{2{\nu}z_0 F(1-\frac{\nu}{2};\frac{3}{2};2{\nu}z_0^2) +
G F(\frac{1-{\nu}}{2};\frac{1}{2};2{\nu}z_0^2)} \right)
\]
and
$z_0 \equiv E/(mc^2)$. The matching condition at $x_1=-a$ for (4.7b)
and
\ren
\begin{equation}
{\psi}^{-}_{even} = s
\left( \begin{array}{c}
mc^2\\
E - i {\kappa}
\end{array} \right)
e^{\frac{1}{{\hbar}c} {\kappa}x_1} ,
\qquad x_1 \leq -a,
\end{equation}
gives the same spectrum equation.

In a similar way we can derive the equation that determines the spectrum
of the odd bound states:
\ren
\begin{equation}
E = mc^2 \cos \left({\lambda}_{\bar{G}}(E) \right).
\end{equation}

\vspace{5 mm}

{\bf 2.} With the self-field term $\bar{A}_{\mu}^{self}$ the Dirac
equation is nonlinear in ${\psi}$, so the spectrum problem becomes
essentially more complicated. As in [2-4], we can solve the problem
in the approximation when the self-interaction contribution to the
energy spectrum is very small with respect to the contribution of
the interaction between the Dirac particle and external field.

Let us assume that the equations (4.10) and (4.12) have $N_{+}$
and $N_{-}$ solutions , correspondingly, i.e. for
$\bar{A}_{\mu}^{self}=0$ there are $N_{+}$ even and $N_{-}$ odd
bound states. The total number of discrete states in the band
$|E|<mc^2$ is then $N=N_{+} +N_{-}$. Let us denote the normalized
bound state eigenfunctions by ${\psi}_n^{ext}$, $n=\overline{1,N}$.
Up to a normalization factor ${\psi}_n^{ext}$ coincide with
(4.7a-b) for even and with (4.8a-b) for odd states in the region
$-a \leq x_1 \leq a$.

In $(3+1)$-dimensions, the self-field effects are of order of the fine
structure constant and higher. This allows us to restrict our calculations
to the first order of this constant. In our study, to make the
self-field effects small we assume that $|e| \ll q$, so the
self-interaction shifts the bound state energies by a small amount
\ren
\begin{equation}
E_n = E_n^{ext} + {\Delta}E_n^{self}
\end{equation}
and does not change the number of states. In (4.13) $E_n^{ext}$
are discrete spectrum energies of the external field problem
without $\bar{A}_{\mu}^{self}$. The eigenfunctions of the discrete
spectrum cannot be characterized now by a definite parity, because
the self-interaction term in the Dirac Hamiltonian is not in general
invariant under parity reversal.

For the nonlinear Dirac equation the superposition principle does
not hold. It is not possible to expand the exact solutions of this
nonlinear equation as a superposition of the linear equation solutions
${\psi}_n^{ext}$ with arbitrary time-dependent coefficients, even
though ${\psi}_n^{ext}$ form a complete set. What can we use now is
the Fourier expansion in the time coordinate [2-4]
\[
{\psi}(x_1,t) = \sum_{n=1}^{N} {\psi}_n(x_1) e^{-\frac{i}{\hbar}
E_nt} + \int_{- \infty}^{- mc^2} dE {\psi}(x_1,E)
e^{- \frac{i}{\hbar} Et}
\]
\ren
\begin{equation}
+ \int_{mc^2}^{\infty} dE {\psi}(x_1,E) e^{-\frac{i}{\hbar} Et}
\end{equation}
in which the time behaviour is known, and of the form
$\exp(-\frac{i}{\hbar}Et)$. The functions ${\psi}_n(x_1)$,
${\psi}(x_1,E)$ and the energies $E_n$ are unknown.

To derive an information about the spectrum it is simpler and more
convenient to work with the action rather than with the Dirac
equation. Since manipulations which we are going to do below are
valid in both discrete and continuous spectrums, we can use
instead of (4.14) the following compact expression
\ren
\begin{equation}
{\psi}(x_1,t) = \overline{\sum}_n {\psi}_n(x_1)
e^{-\frac{i}{\hbar}E_nt},
\end{equation}
where $\overline{\sum}_n$ means summation over discrete states
and integration over continuous ones.

If we insert the Fourier expansion into the action, then up to
the terms depending only on the external field we obtain
\[
S_0 = \overline{\sum}_{n,m} \int_{-\infty}^{\infty} dt
\int_{-\infty}^{\infty} dx_1 \bar{\psi}_n(x_1)
\left[ {\gamma}^{\mu}(i{\hbar}c{\partial}_{\mu}
- e\bar{A}_{\mu}^{ext}) -mc^2 \right] {\psi}_m(x_1)
e^{\frac{i}{\hbar}{\omega}_{nm}t},
\]
where ${\omega}_{nm} \equiv E_n -E_m$, and
\[
S_{self} = -\frac{e^2c}{2} \overline{\sum}_{n,m,r,s}
\int_{-\infty}^{\infty} dt \int_{-\infty}^{\infty} dx_1
\int_{-\infty}^{\infty} dt^{\prime} \int_{-\infty}^{\infty} dx_1^{\prime}
[ j_{nm}^{\mu}(x_1) D^c(x_1-x_1^{\prime};t-t^{\prime})
j_{rs,{\mu}}(x_1^{\prime})
\]
\[
- j_{nm}^0(x_1) D^c(0;t-t^{\prime}) j_{rs,0}(x_1^{\prime}) ]
e^{\frac{i}{\hbar} ({\omega}_{nm} t + {\omega}_{rs} t^{\prime})},
\]
where
\[
j_{nm}^{\mu}(x_1) \equiv \bar{\psi}_n(x_1) {\gamma}^{\mu} {\psi}_m(x_1).
\]
After time integration for $S_0$ we find
\ren
\begin{equation}
S_0 = 2{\pi}{\hbar} \overline{\sum}_{n,m} \int_{-\infty}^{\infty}
dx_1 {\psi}^{\star}_n(x_1) (E_m-H) {\psi}_m(x_1) \delta({\omega}_{nm}),
\end{equation}
with $H$ given by (2.5) . If the
${\psi}_n(x_1)$ were solutions of the external field problem with
the vanishing self-potential $\bar{A}_{\mu}^{self}$, i.e.
${\psi}_n^{ext}(x_1)$, then this expression would be zero for
$E_n = E_n^{ext}$. The ${\psi}_n^{ext}(x_1)$ minimize $S_0$
alone. However, now the entire action $S$ , of which $S_0$ is
only one term, must be minimized as a whole.

Time integrations in $S_{self}$ can be performed using (3.6) and
we can write the self-field part of the action entirely in terms of the
Fourier components of the currents
\ren
\begin{equation}
S_{self} = \frac{e^2{\hbar}^2}{2} \overline{\sum}_{n,m,r,s}
\delta({\omega}_{nm} + {\omega}_{rs}) \int_{-\infty}^{\infty}
\frac{dq_1}{\frac{1}{c}{\omega}_{nm}^2 -q_1^2 +i\epsilon} \cdot
(T_{nm}^{\mu}(q_1) T_{rs,{\mu}}(-q_1) - T_{nm}^0(0) T_{rs}^0(0) ),
\end{equation}
where
\ren
\begin{equation}
T_{nm}^{\mu}(q_1) \equiv \int_{-\infty}^{\infty} dx_1
j_{nm}^{\mu}(x_1) e^{\frac{i}{\hbar}q_1x_1} .
\end{equation}
The ${\psi}_n(x_1)$ here are still exact solutions for which the
action $S$ will vanish identically.

In the approximation of small self-interaction contribution, we can
solve the spectrum problem iteratively. To lowest order of iteration
we replace ${\psi}_n(x_1)$ with ${\psi}_n^{ext}(x_1)$ and then
solve for $E_n$ which have the form (4.13). The number of discrete
states without and with the self-field term $\bar{A}_{\mu}^{self}$
is assumed to be the same, and transitions between discrete and
continuous states are neglected. Omitting the integration over continuous
states and using the orthonormality of ${\psi}_n^{ext}$, we write
$S_0$ in the form
\ren
\begin{equation}
S_0 = 2{\pi}{\hbar} \sum_{n,m=1}^{N} {\Delta}E_n^{self}
\cdot \delta({\omega}_{nm}) {\delta}_{nm}.
\end{equation}

In the self-field part of the action we separate the terms
according to $E_n=E_m$, $E_r=E_s$ and according to $E_s=E_n$,
$E_r=E_m$, the only two ways of satisfying the overall
$\delta$-function for discrete spectrum energies. And since
$S=0$ to this order of iteration we can solve for ${\Delta}E_n^{self}$.
The action and the total energy of the system are related by a
$\delta$-function. Cancelling this $\delta$-function as well as
the sum over $n$ to obtain the energy shift of a fixed level $n$,
we get
\[
{\Delta}E_n^{self} = \frac{e^2{\hbar}}{4{\pi}} \sum_{m=1}^{N}
\int_{-\infty}^{\infty} dq_1 {\cal P}\frac{1}{q_1^2} \cdot
G_{nn,mm}(q_1) - \frac{e^2{\hbar}}{8{\pi}} \sum_{m=1}^{N}
\int_{-\infty}^{\infty} \frac{dq_1}{|q_1|} {\cal P}
\frac{1}{\frac{1}{c}{\omega}_{nm} - |q_1|} (G_{nm,mn}(q_1)
\]
\ren
\begin{equation}
+ G_{mn,nm}(q_1))
+ i \frac{e^2{\hbar}c}{4} \sum_{(m<n)} \frac{1}{{\omega}_{nm}}
{\rm Re}\left[ G_{nm,mn}(\frac{1}{c}{\omega}_{nm}) +
G_{nm,mn}(\frac{1}{c}{\omega}_{mn}) \right],
\end{equation}
where
\[
G_{nm,rs}(q_1) \equiv T^{ext,{\mu}}_{nm}(q_1) T^{ext}_{rs,{\mu}}
(-q_1) - T^{ext,0}_{nm}(0) T_{rs}^{ext,0}(0),
\]
${\cal P}$ stands for the principal value prescription, while
${\rm Re}$ means the real part. The subscript ext here indicates
that the Fourier components (4.18) must be calculated by using
${\psi}_n^{ext}$.

The last term in (4.20) contributes if $m<n$. This shows that
only the ground state $n=1$ of our $N$-level system is stable.
The energy shift for the excited states $n=2, . . .,N$ is
complex, so the imaginary part of the shift can be identified with
the spontaneous emission of the excited states due to self-interaction.

\section{Conclusions}

{\bf 1.} We have studied the bound state problem for the Dirac
particle moving in both external and its own radiation fields
in $(1+1)$-dimensions. We have shown that if the total electric
charge of the system vanishes, then the asymptotically linearly
rising part of the external potential which was responsible for
nonexistence of bound states in the external field problem
without self-interaction is cancelled by the self-potential
of the zero mode of the Dirac particle charge density. The
resulting Dirac equation includes only that part of the external
potential which is finite at spatial infinities and also the
self-potential created by the non-zero modes of the charge density.

We have proved that this equation has a set of solutions which
show that the Dirac particle and external charges are confined
in a stable system. These solutions correspond to energy levels
in the band $|E|<mc^2$. Only the lowest level is precise. The
higher levels have a nonzero linewidth which manifests itself
as spontaneous emission. This picture is characteristic for
hydrogenlike atoms. According to the self-field approach, the
hydrogen atom has no precisely defined sharp energy levels,
other than the ground state [2-4]. The excited states cannot be
stable due to radiation reaction.

{\bf 2.} We have solved the bound state problem for the external
potential created by two positive static charges of the same value.
All other cases when we have two or more external charges can be
considered analogously.

The exceptional case is one external charge at the origin. The potential
created by this charge consists only of the asymptotically linearly
rising part, so the corresponding nonlinear Dirac equation does
not include any external field. This means that the approximation
when the self-interaction contribution to the spectrum
is small with respect to the one of the external field cannot be
applied. We need here to look for other ways to solve the
nonlinear Dirac equation. This problem remains open.

{\bf 3.} Vanishing of the total electric charge is the condition
necessary for the existence of bound states of the Dirac particle
and external static charges on line. If the total electric charge
is not zero, then the Dirac equation does not reveal bound state
solutions. Therefore, in the Dirac theory on line bound states
can be only neutral. This is one of the pecularities of
one-dimensional physics.

We can consider charged states as well, but these states are not
stable. The presence of an uncompensated electric charge leads
to the linearly rising Coulomb potential in the Dirac equation
and to a potential of the inverted oscillator type in the
corresponding Schr\"odinger equation. Such potentials are
known to allow only metastable states [12]. So for limited times
the Dirac particles and external charges can be confined in
a charged metastable state. Metastable states can decay into
stable ones. If particles emitted in decay take away a charge
equal to the charge of the metastable state, then the
remaining particles can create a neutral bound state.

\newpage

\begin{center}

{\bf Figure Captions}

\end{center}

\begin{flushleft}

\vspace{5 mm}

Figure $1$ \\

\vspace{2 mm}

The potential $\bar{A}_0(x_1)$ in the case of two external
charges of the same value $q>0$ at the points $x_1=\pm a$.

\end{flushleft}

\newpage

\setlength{\unitlength}{1cm}
\begin{center}
\begin{picture}(12,10)(-6,0)

\put (0,5){\line(0,1){3}}
\put (0,8){\line(0,1){1}}
\put (0,9){\vector(0,1){1}}

\put (0,5){\line(0,-1){3}}
\put (0,2){\line(0,-1){2}}

{\thicklines
\put (-6,5){\line(1,0){2}}
\put (4,5){\line(1,0){2}}}

\put (-4,5){\line(1,0){1}}
\put (-3,5){\line(1,0){3}}
\put (3,5){\line(1,0){1}}
\put (5,5){\vector(1,0){1}}
\put (0,5){\line(1,0){3}}

\put (6,5.2){$x_{1}$}
\put (.2,10){$\bar{A}_0^{ext}$}

{\thicklines
\put (0,2){\line(4,3){4}}
\put (0,2){\line(-4,3){4}}}

\put (0.3,4.8){\makebox(0,0)[tr]{0}}
\put (4.2,4.8){\makebox(0,0)[tr]{a}}
\put (-4.2,4.8){\makebox(0,0)[tl]{-a}}
\put (0.7,1.8){\makebox(0,0)[tr]{-qa}}

\end{picture}

\vspace{2 cm}

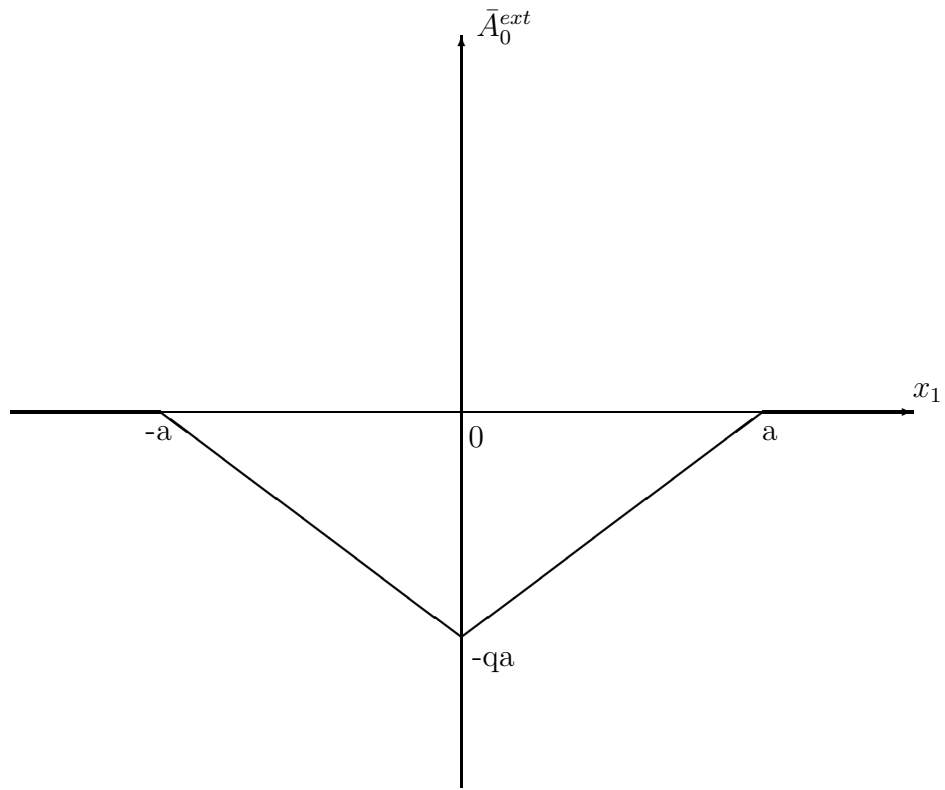
{\bf FIG. 1.}

\end{center}


\newpage

\begin{thebibliography}{99}
\bibitem{ga} Gali\'c H 1988 Am.J.Phys. {\bf 56} 312
\bibitem{ba} Barut A O and Kraus J 1983 Found.Phys. {\bf 13} 189
\bibitem{rut} Barut A O 1990 {\it New Frontiers in Quantum
Electrodynamics and Quantum Optics} ed A O Barut (New York: Plenum)
\bibitem{dow} Barut A O and Dowling J P 1990 Phys.Rev. {\bf A41} 2284
\bibitem{dir} Dirac P A M 1958 {\it The Prinsiples of Quantum
Mechanics} (Oxford: Clarendon)
\bibitem{me} Messiah A 1962 {\it Quantum Mechanics} vol.2 (Amsterdam:
North Holland)
\bibitem{bar} Barton G 1986 Ann.Phys. NY {\bf 166} 322
\bibitem{sch} Schwinger J 1962 Phys.Rev. {\bf D128} 2425\\
Schwinger J 1963 {\it Theoretical Physics, Trieste Lectures}
(Vienna: IAEA)
\bibitem{mant} Manton N S 1985 Ann.Phys. NY {\bf 159} 220
\bibitem{iso} Iso S and Murayama H 1990 Progr.Theor.Phys. {\bf 84}
142
\bibitem{han} Hanson A J, Peccei R D, and Prasad M K 1977
Nucl.Phys. {\bf B121} 477
\bibitem{nor} Dombey N and Saradzhev F M 2000 J.Phys. {\bf A33} 4491
\end{thebibliography}
\end{document}